# Flipping growth orientation of nanographitic structures by plasma enhanced chemical vapor deposition

Subrata Ghosh*, K. Ganesan*, S. R. Polaki, S. Ilango, S. Amirthapandian, S. Dhara, M. Kamruddin, and A. K. Tyagi

*Materials Science Group, Indira Gandhi Centre for Atomic Research, Kalpakkam - 603102, India.
* These authors contributed equally to this work. Email: kganesan@igcar.gov.in*

Abstract

Nanographitic structures (NGSs) with multitude of morphological features are grown on $SiO_2$/Si substrates by electron cyclotron resonance - plasma enhanced chemical vapor deposition (ECR-PECVD). $CH_4$ is used as source gas with Ar and $H_2$ as dilutants. Field emission scanning electron microscopy, high resolution transmission electron microscopy (HRTEM) and Raman spectroscopy are used to study the structural and morphological features of the grown films. Herein, we demonstrate, how the morphology can be tuned from planar to vertical structure using single control parameter namely, dilution of $CH_4$ with Ar and/or $H_2$. Our results show that the competitive growth and etching processes dictate the morphology of the NGSs. While Ar-rich composition favors vertically oriented graphene nanosheets, $H_2$-rich composition aids growth of planar films. Raman analysis reveals dilution of $CH_4$ with either Ar or $H_2$ or in combination helps to improve the structural quality of the films. Line shape analysis of Raman 2D band shows nearly symmetric Lorentzian profile which confirms the turbostratic nature of the grown NGSs. Further, this aspect is elucidated by HRTEM studies by observing elliptical diffraction pattern. Based on these experiments, a comprehensive understanding is obtained on the growth and structural properties of NGSs grown over a wide range of feedstock compositions.

## Introduction

The carbon nanostructures such as fullerenes, carbon nanotubes (CNTs), graphene and nanodiamonds have drawn a significant attention in scientific and industrial research due to their unique structural, electrical, optical and mechanical properties.[1-4] Recently, a considerable interest is shown to grow large area vertical graphene nanosheets (VGNs), also called as carbon nanowalls (CNWs), and a few layer planar nanographite (FPNG) structures because of their potential applications in various fields.[2-5] The VGNs have unique structural and electrical characteristics with high aspect ratio and large three dimensional networks which make them an excellent candidate for field emission, fuel cells, chemical & biosensors and energy storage devices.[2-6] The FPNG films can also be used for transparent conductive films and flexible electronics.[5]

Among a variety of growth processes of nanographite structures (NGSs), the plasma enhanced chemical vapor deposition (PECVD) holds the promise to grow large area, conformal and uniform NGSs in a controlled manner with numerous advantages. Also, it is important to note that the synthesis is catalyst free and it is possible to grow on any type of substrates such as metals, semiconductors or insulators. Thus, the PECVD growth helps to avoid a major hurdle of transferring the graphene onto a dielectric substrate for device fabrication. The growth parameters such as feedstock gases and their composition, microwave power, electric field, temperature and pressure have strong influence on the morphology, growth rate and structural properties of the NGSs.[7-9] In order to realize and enhance the full potential of this material, it is necessary to have a clear understanding and fine control of their structural and physical properties by tuning the growth conditions.

Although a number of reports have appeared in literature on this subject describing the role of different gas compositions on the growth of NGSs using a variety of growth techniques, a systematic study using a single growth technique is scarce.[10-12] Further, the observed results in terms of morphology and structural properties of NGSs grown by various groups are only in partial agreement, and sometimes even contradictory.[3, 5-27] The reason for such variation and discrepancy among the published results could be due to different synthesis techniques and conditions, which make it difficult to give a unique interpretation for the growth and its properties. Hence, a systematic study on growth of NGSs under different gas compositions using a single growth technique is highly desirable. Also, this could possibly explain the structure-property relationship of NGSs which is essential for using them in applications. A brief survey on the existing reports on the growth of VGNs and FPNG films in this context as follows:

In 2002, Wu *et al.*[13] incidentally observed the growth of vertically oriented well separated graphene sheets, called as CNWs, while synthesizing CNTs by microwave PECVD using metallic catalysts. Since then, several groups have explored the growth of VGNs on catalytic as well as non-catalytic metal / dielectric surfaces using various plasma based techniques such as microwave PECVD, dc plasma discharge, ICP-plasma and thermal plasma jet systems.[10-27] In an excellent review, Bo *et al.*[11] summarized the influence of various key process parameters on the synthesis and growth model for NGSs on various types of substrates. In spite of several existing reports, herein we note that, still there is no unique theory that could unveil the atomistic growth mechanism and give a prescription to optimize the process parameters for a given plasma source.[11] In short, while some of the groups have used Ar for the dilution of hydrocarbon,[23-25] the others employed $H_2$ to grow VGNs.[7-10, 13, 26] A few groups had also used combination of Ar and $H_2$ for the growth of VGNs.[20, 21]

The growth of monolayer graphene or FPNG films on metal surfaces using PECVD have been successfully demonstrated by a few groups.[5, 14-18] A single layer graphene on Cu or Ni are grown using $CH_4/H_2$ or $CH_4/Ar$ by microwave PECVD.[14, 16, 17] A high quality single layer graphene on copper substrate is grown by ICP-CVD using feedstock gases of $CH_4:Ar:H_2$ in 1:90:10 ratio.[18] Invariably all the above groups have used either $H_2$ and/or Ar as the gas for dilution to grow planar structures.

In addition to growth of NGSs, Cho *et al.*[19] had controlled the density of CNWs deposited using $CH_4/H_2$ at 2:1 ratio by varying total pressure and discharge power during growth in radical injection PECVD. Apart from single carrier gas, by varying the composition of two different types of dilution gases like Ar and $H_2$, the possibility to grow tree-like or petal-like structures is also reported.[20] Ubnoske *et al.*[22] could change the morphology of carbon nanostructures from CNTs to vertically oriented graphene nanosheets with intermediate graphenated CNTs by tuning the substrate temperature from 800 to 1100 $^0$C in microwave PECVD with $CH_4:NH_3$ flow in ratio of 3:1 sccm. Thus, only a limited study has been carried out on the nature of tunable morphology and their structural properties.

In this work, we report a systematic study on the synthesis of NGSs with tunable morphology using electron cyclotron resonance (ECR) - PECVD. We also try to address a number of issues relating to the effect of dilution gases, Ar and $H_2$, on the growth and structural properties of NGSs deposited on dielectric $SiO_2$/Si substrates. Morphology is tuned from vertical to planar NGSs by the appropriate choice of carrier gas composition. A plausible growth mechanism on different morphological features is discussed. The structural properties of the grown films are analyzed in the light of Raman spectroscopy. It is found that the data obtained are in good agreement with morphology obtained from field emission scanning electron microscopy (SEM).

**Experimental**

**Growth of NGSs by ECR-PECVD**

The synthesis of NGSs on $SiO_2$/Si substrates is performed under different feedstock gas compositions using ECR-PECVD. The details of the deposition system is reported elsewhere.[23] For the present study, we choose $CH_4$ (5N purity) as hydrocarbon source. The Ar (3N purity) and $H_2$ (5N purity) are used as dilution gases. Thermally oxidized $SiO_2$ substrates of thickness about 300 nm are kept at 5 and 40 cms below the gas shower head and quartz window, respectively. Prior to growth, the chamber is evacuated down to a base pressure of $5\times10^{-6}$ mbar by a turbomolecular pump and then the substrate temperature is increased to $800^oC$ at a rate of $30^oC$/min and maintained for 30 min. Subsequently, the substrates are pre-cleaned by Ar plasma with 20 sccm flow for 10 min at 200W power. After cleaning, $CH_4$ is fed into the chamber through downstream gas shower ring along with the dilution gas(es) $H_2$ or Ar or $H_2$-Ar combination. For all growth experiments, the microwave power is kept constant at 400 W. After the growth of NGSs (30 minutes), the plasma is turned off and the samples are allowed to remain at growth temperature for another 30 minutes. Samples are then cooled to room temperature by switching OFF the heater.

The samples are investigated for their morphological and structural features by SEM, atomic force microscope (AFM) and high resolution transmission electron microscopy (HRTEM, LIBRA 200FE, Zeiss). HRTEM samples are prepared by scratching VGNs grown on $SiO_2$ substrates by Cu grids coated with lacey carbon. The degree of graphitization, in terms of defects and disorder, are evaluated by micro-Raman spectroscopy (inVia Renishaw) in the back scattering geometry. The spectrometer is equipped with $Ar^+$ laser (514 nm) as the probe source with grating monochromator (1800 grooves/mm) and Peltier cooled charged couple device as detector. A microscope objective of 100X magnification with numerical aperture of 0.8 is used in this study. In order to avoid the laser induced heating on samples, laser power is kept below 1 mW. Raman spectra are analyzed using WIRE3.2 software and fitted with Lorentzian function.

## Results

The growth experiments were carried out at various input gas compositions having different $CH_4:Ar:H_2$ ratios while the other parameters like growth time, temperature, microwave power and post-annealing time are kept constant. To make the analysis simple, we divide our experimental results into three cases according to the type of dilution viz (i) dilution of $CH_4$ with Ar (case 1), (ii) dilution of $CH_4$ with $H_2$ (case 2) and (iii) dilution of $CH_4$ with a mixture of Ar and $H_2$ (case 3). In order to study the effect of dilution, we retain the hydrocarbon content to be constant in all the growth experiments except for case 3. Hence, the total pressure in the chamber is maintained between $1.7 \times 10^{-3}$ and $5.5 \times 10^{-3}$ mbar for the total flow rate which changes from 5 to 40 sccm. All the samples are labeled in accordance with this gas ratios in the order $CH_4:Ar:H_2$ unless or otherwise mentioned specifically.

## SEM Analysis

## Case 1: Dilution of $CH_4$ with Ar

Fig. 1 shows the surface morphology of the NGSs grown with $CH_4$ diluted with Ar in the ratio of $CH_4:Ar$ upto 1:7. It can be seen from Fig. 1(a) that mere decomposition of $CH_4$ alone can produce a homogeneous film consisting of many island like structures over the smooth substrate. Subsequently when Ar is introduced into $CH_4$, the nanoisland structures get transformed into vertically oriented structure as shown in Fig. 1(b). Further increase in Ar, increases the height of the VGNs and decreases the number density of the sheets. However, the width and length of the vertical sheets increase with Ar dilution as shown in Fig. 1(a-e). In addition, substantial increase in height from 27.8 to 164.9 nm is observed when the Ar flow is increased from 5 to 35 sccm (see Fig. 1(f)). The in-built electric field associated with plasma is found to assist vertical growth.[13] Further, the plasma can produce a relatively higher temperature and chemical potential gradients near the substrate surface which can also enhance the vertical growth.[28] The dilution of hydrocarbon with Ar is known to enhance the formation of $C_2$ radicals which helps to grow high quality vertical graphitic structures at higher growth rate.[9, 24, 25] However, as the growth progresses, the atomic hydrogen ($H^*$) generated from the dissociation of $CH_4$ by microwave energy also start etching the growing film. The concurrent growth and etching processes make the VGNs to gain height and decrease the number density.

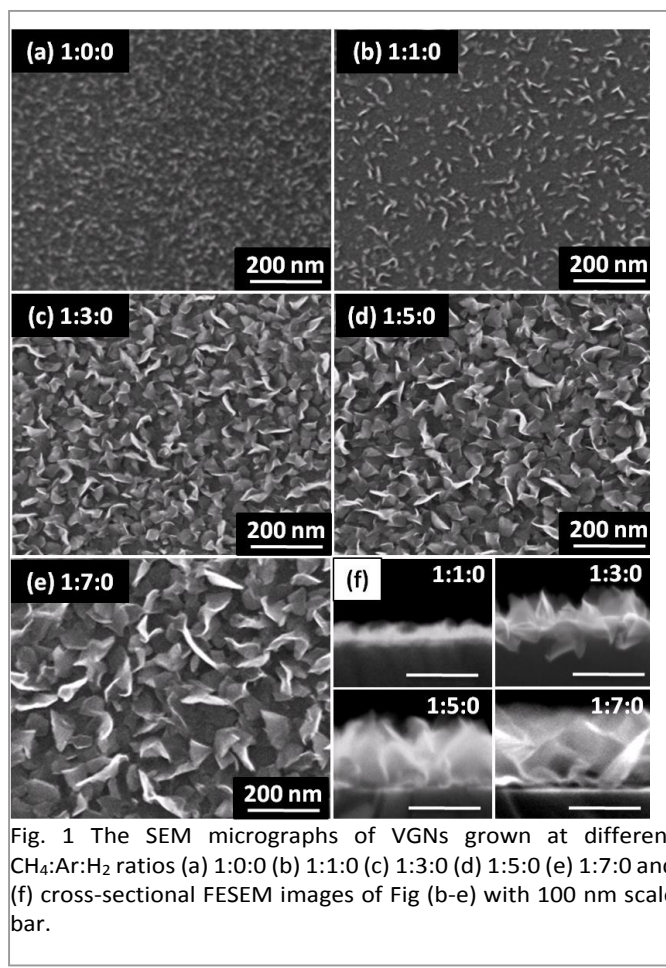

Fig. 1 The SEM micrographs of VGNs grown at different $CH_4$:Ar:$H_2$ ratios (a) 1:0:0 (b) 1:1:0 (c) 1:3:0 (d) 1:5:0 (e) 1:7:0 and (f) cross-sectional FESEM images of Fig (b-e) with 100 nm scale bar.

## Case 2: Dilution of $CH_4$ with $H_2$

Fig. 2 shows the surface morphology of the NGSs grown with $CH_4$ diluted with $H_2$ in the ratio of $CH_4$:$H_2$ upto 1:7..The SEM images show no specific features and the surface appears to be smooth. However, we perform AFM measurements to have a closer look and access the fine features of the surface and shown in Fig. 3. When $H_2$ is introduced into $CH_4$, i.e. 1:1 and 1:3 ratios of $CH_4$: $H_2$, the nanoisland structure (see Fig. 2(a)) turned out to be a smooth planar structure (see Fig. 2(b) and (c)). Further increase in $H_2$ flow leads to the formation of nanoisland structure over electrically continuous NGSs, and a discontinuous film for flow rates of $CH_4$:$H_2$ at 1:5 and 1:7 $H_2$ plays an important role in decomposition kinetics of $CH_4$ under microwave plasma. $C_2$ and $H^*$ radical density can be effectively controlled by $H_2$ content in the feedstock gases. Further, the $H^*$ radicals efficiently etch out amorphous carbon (a-C). It also etches out weakly bonded carbon atoms and small graphitic fragments to prevent formation of additional graphene layers. Further, it enables the mobility of carbon adatoms to efficiently reduce the structural disorder.[6]

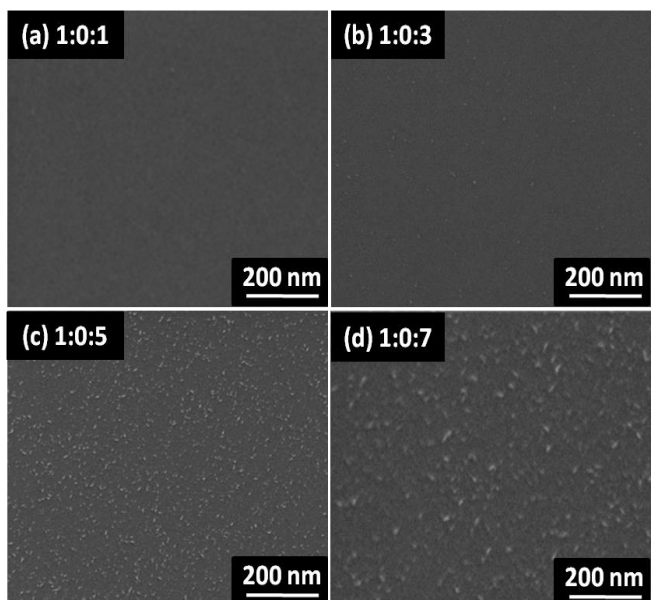

Fig. 2 FESEM micrographs of a few layer planar nanographite films grown at different $CH_4:Ar:H_2$ ratios
(a) 1:0:1 (b) 1:0:3 (c) 1:0:5 and (d) 1:0:7. Smooth and planar films are observed when the gas flow
ratios are 1:0:1 and 1:0:3 (fig 2a & 2b). Isolated clusters over electrically continuous nanographitic
film are observed for 1:0:5 flow ratio (fig 2c). Discrete nanographitic clusters over $SiO_2$ surface is
observed for 1:0:7 flow ratio (fig 2d).

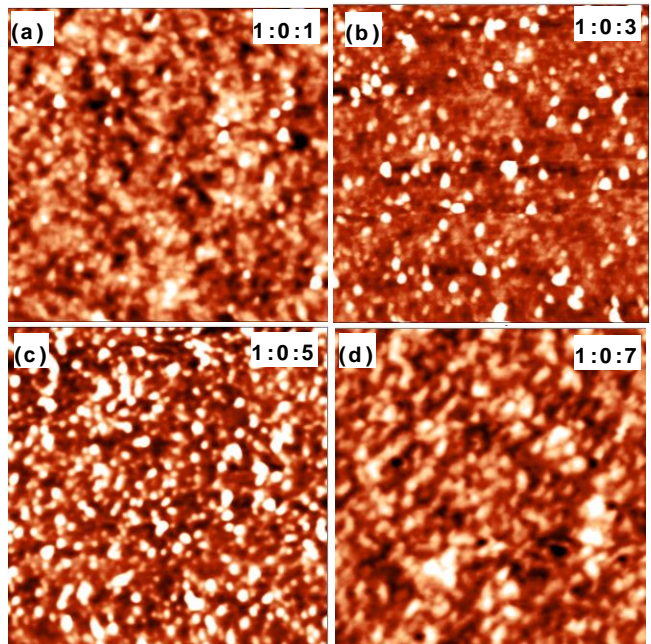

Fig. 3. AFM topography of the films grown at dilution of $CH_4$ with $H_2$ at the ratio of (a) 1:1 (b) 1:3, (c)
1:5 and (d) 1:7. Each image is recorded over an area of 500 x 500 $nm^2$. The peak to peak height variation
is found to be maximum of 10.1, 14.8, 15.7 and 7.8 nm for the films grown at 1:1, 1:3, 1:5 and 1:7 ratios
respectively. The z-scale is adjusted appropriately in order to have maximum visibility of the ratios, respectively. Thus, with increase in $H_2$ content, discontinuity in the films is observed with isolated clusters, and the growth of the film is hindered as shown in Fig. 2(a-d).

Fig. 3 shows the topographic images obtained from AFM. These images clearly show the variation on the surface features due to etching of $H^*$ radicals. The rms roughness measured over an area of 500 x 500 $nm^2$ is found to be 0.65, 1.28, 1.57 and 0.82 nm for the films grown at 1:1, 1:3, 1:5 and 1:7 ratios, respectively. The increase in rms roughness values with $H_2$ dilution indicates that the $H^*$ radicals etch out the growing carbon films. At 1:7 dilution ratio, the majority of the carbon films are etched out and only isolated carbon clusters are left out. The thickness of the films is also measured using AFM over a surface scratched by a soft toothpick and the values are given in Table 1. In addition, the ECR plasma produces higher density of $H^*$ radicals in comparison to other types of plasma sources. This excess $H^*$ radicals lead to chemical etching of growing film. Thus, the vertical growth is hampered by excess hydrogen radicals in ECR- PECVD under hydrogen-rich dilution.

## Case 3: Dilution of $CH_4$ with a gas mixture of Ar + $H_2$

As discussed in previous sections (case 1 & 2), the dilution of $CH_4$ with Ar or $H_2$ favors VGNs and FPNG films, respectively. Taking cue from these experiments, a combination of these two dilution gases with $CH_4$ is explored to study their impact on the morphology, growth rate and structural properties. These experiments are performed under three subcategories viz (i) repetition of case 1 experiments with addition of constant $H_2$ (case 3(a)) (ii) repetition of case 2 experiments with addition of constant Ar (case 3(b)) and (iii) increasing $CH_4$ concentration with fixed Ar and $H_2$ ratios (case 3(c)).

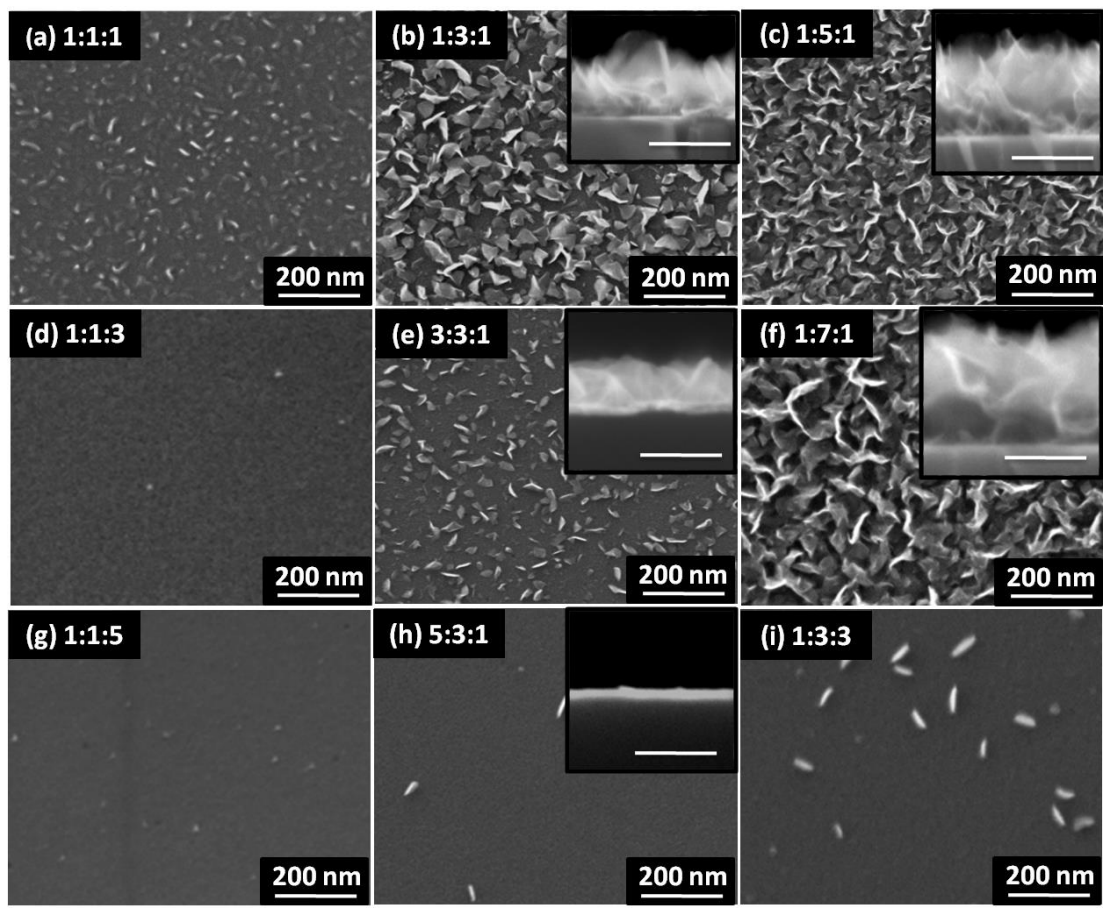

Fig. 4 SEM micrographs of nanographite structures grown under mixed Ar+H$_2$ dilution. The actual gas flow ratios are labeled on each image as CH$_4$:Ar:H$_2$. The inset in Fig (b), (c), (e), (f) and (h) is cross-sectional SEM image of respective film with 200 nm scale bar.

Fig. 4 shows the SEM micrographs for the films grown by dilution of CH$_4$ with a gas mixture of Ar with addition of constant H$_2$ at various concentrations. While repeating the case 1 and case 2 experiments with mixture of constant H$_2$ and Ar, respectively, the morphology of the NGSs is found to change considerably. It is evidenced from these SEM micrographs that the morphology can be tuned from perfect planar to highly dense vertical structures with intermediate sparsely distributed vertical structures which can be seen from in Fig. 4. In case 3(c), the concentration of CH$_4$ alone is increased while maintaining the amounts of Ar and H$_2$ (1:3:1, 3:3:1 and 5:3:1 ratios) constant. The SEM morphology of these films are shown in Fig. 4(b,e,h). As shown in these images, the vertical structure systematically transforms into planar structure with increasing CH$_4$ concentration. The reason for the flipping of orientation is due to the abundance of nascent hydrogen that gets released from fragmenting CH$_4$ in the plasma. Briefly, we observe that the morphology of NGSs is mostly dependant on the nature of dominant dilutant gas. Hence, it is very clear that the surface morphology and aerial density of the nanosheets can be completely controlled using optimal dilution ratios of Ar and $H_2$ in $CH_4$.

**HRTEM analysis**

Fig. 5 (a,b) and (d,e) represent the HRTEM images of the films grown under the flow ratio of 1:7:0 and 1:0:5 respectively. HRTEM analysis clearly confirms that NGSs are composed of about 13 number of graphene layers as shown in Fig 5(a) and 5(b). The graphene sheets are by and large uniform in nature. However, the edges of VGNs are found to be tapered (Fig. 5a). The measured inter layer spacing of the respective samples are found to be 0.356 and 0.364 nm, which is higher than that of graphite (0.334 nm). The increased interlayer spacing could be correlated to the stacking fault disorder called turbostratic graphite.[29]  Fig 5c and 5f show the Fast Fourier transform (FFT) images that are corresponding to the HRTEM images of Fig. 5(b) and Fig.5(e) respectively. Both the FFT images show the spots in elliptical pattern which is a characteristic feature of turbostratic graphite. The enhanced growth kinetics under non-equilibrium conditions in the plasma leads to the turbostratic disorder.

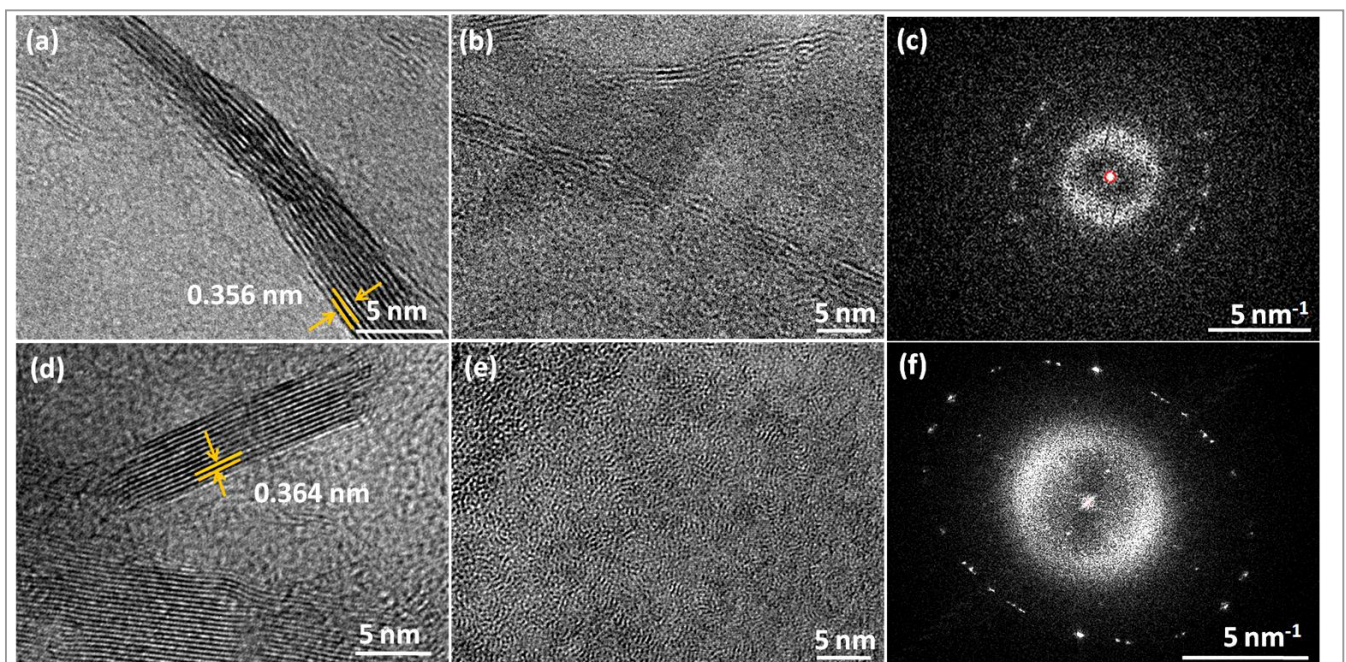

Fig. 5(a,b) and 5(d,e) are the HRTEM images  of the samples grown under feedstock composition of 1:7:0 and 1:0:5 respectively. Fig 5c and 5f are fast Fourier transform images corresponding to image (b) and (e) respectively.

## Raman spectroscopy

Raman spectroscopy is a well established and powerful technique for structural characterization of carbon based materials. We employed this technique to characterize the VGNs and FPNGs. Fig. 6 shows the Raman spectra of the films grown at different CH4, Ar and H2 compositions. The Raman spectra of the samples are fitted with Lorentzian profile and the calculated parameters are given in Table 1. As shown in Fig. 6, the spectra consist of several Raman modes such as D, G, D′, G′ (also called as 2D) and D+D′ bands which are typical for a defective graphitic system.[23],[27] The G band originates from in-plane vibration of sp2 carbon atoms with E2g symmetry at the Brillouin zone center. The D and D′ bands, respectively originate from the process of inter-valley and intra-valley double resonance which are active only in defective graphitic system. The 2D′ band is the overtone of D′ band. The peak at ~ 2485 $cm^{-1}$ is assigned to D+D′′ band which is due to the combination of D phonon and the zone boundary phonon corresponding to LA branch. The weak and broad band at around 1100 $cm^{-1}$ (D′′ band) which can arise due to high density of edges, bond stretching mode of $sp^3$ hydrogenated carbon and presence of pentagon rings.[30] Due to polarized Raman effect, $I_D/I_G$ ratio is always large at edges than the basal plane. On the other hand, FWHM of G band does not significantly vary at edges except for local doping due to edge functionalisation.[31] Since the vertical structures have extremely high density of graphene edges and the FPNG do not have any edges except grain boundaries, we adopted FWHM rather than established $I_D/I_G$ ratio for structural analysis of the NGSs.

The sample, grown without any dilution gas (Curve no - 1:0:0 in Fig. 5a), exhibits complete overlapping of G and D′ bands and the observed higher full width at half maximum (FWHM) of G′ band ( ~ 109 $cm^{-1}$ ) indicates defect concentration is high in this sample. As in the case 1, the addition of CH4 with Ar leads to continuous decrease in D′ band intensity and make it appear as a shoulder in G band (Fig. 6(a)). This indicates a reduction of defects with increasing Ar content. Further, the case 3(a), also exhibits similar trend in terms of defects as evidenced from Fig. 6(b). For case 1 (Table 1), the FWHM of D, G & G' bands decrease monotonically with Ar dilution that confirms the reduction in the defect concentration. On the other hand, for case 3(a), no systematic variation is observed when a constant H2 added along with Ar.

Fig. 6(c) shows the Raman spectra for the samples grown by dilution of $CH_4$ with $H_2$ which is assigned as case 2. The FWHM of the D, G and G' bands decrease monotonically with $H_2$ dilution as shown in the Table 1 except for higher $H_2$ concentration (1:0:$\geq$5). Also, it clearly shows that there is a decrease in disorder as evidenced from the splitting of G and D' bands in contrast to the film grown without dilution. However, the defect density is found to be still larger than case 1. On the other hand, in case of 3(b) as shown in Fig. 6(d), the defect density is reduced further when Ar is introduced and is also evidenced from Table 1. Fig. 6(e) shows the Raman spectra for samples grown at increasing concentration of $CH_4$ with fixed Ar and $H_2$ content. We observe the D' band intensity to decrease continuously with $CH_4$ and appears as a small hump in the G band.

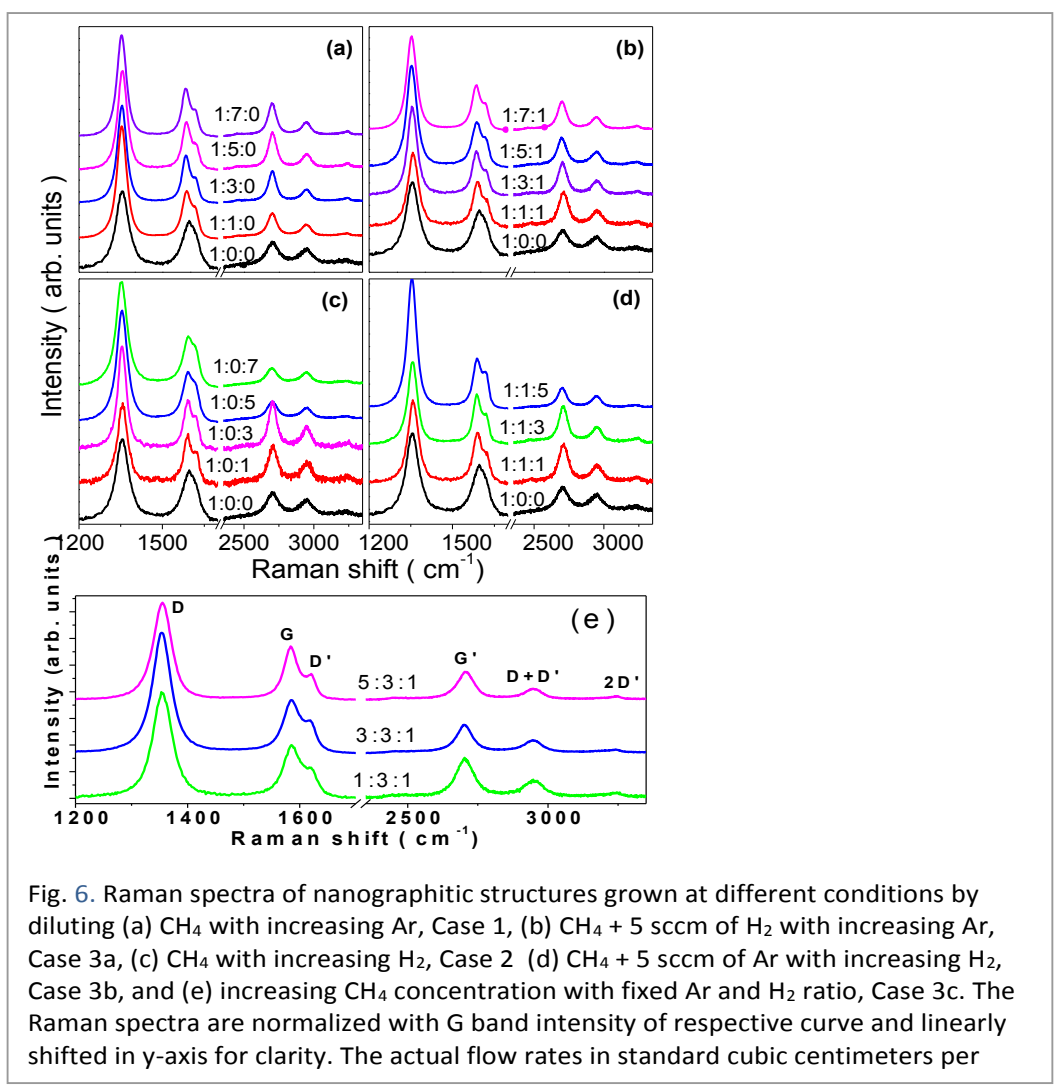

Fig. 6. Raman spectra of nanographitic structures grown at different conditions by diluting (a) $CH_4$ with increasing Ar, Case 1, (b) $CH_4$ + 5 sccm of $H_2$ with increasing Ar, Case 3a, (c) $CH_4$ with increasing $H_2$, Case 2 (d) $CH_4$ + 5 sccm of Ar with increasing $H_2$, Case 3b, and (e) increasing $CH_4$ concentration with fixed Ar and $H_2$ ratio, Case 3c. The Raman spectra are normalized with G band intensity of respective curve and linearly shifted in y-axis for clarity. The actual flow rates in standard cubic centimeters per

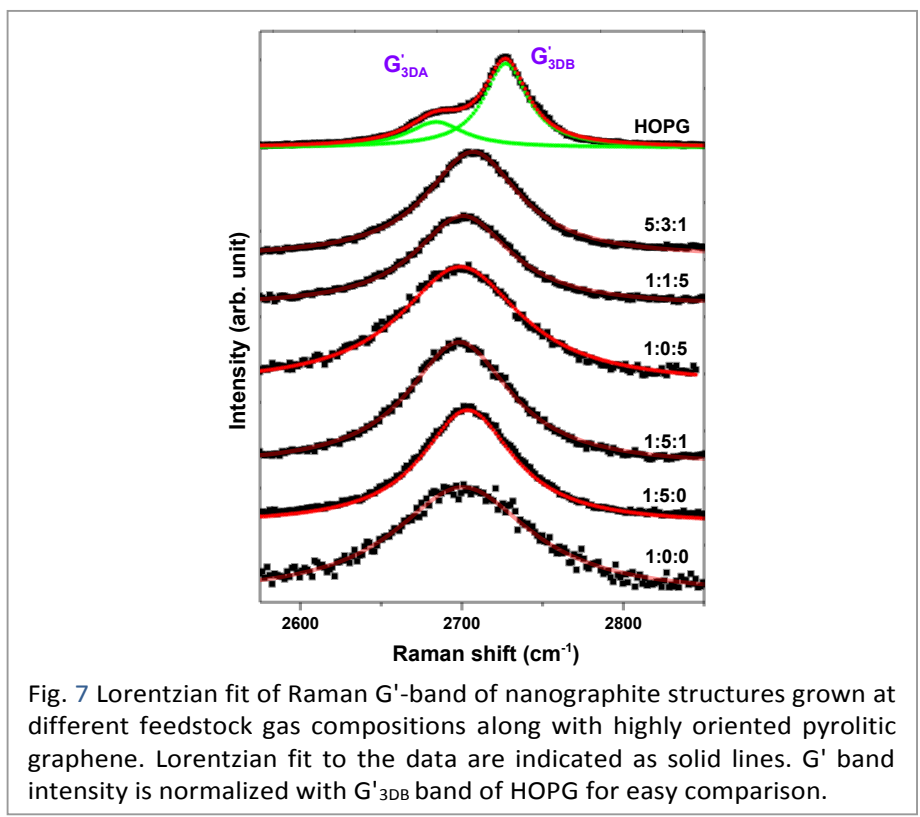

Fig. 7 Lorentzian fit of Raman G'-band of nanographite structures grown at different feedstock gas compositions along with highly oriented pyrolitic graphene. Lorentzian fit to the data are indicated as solid lines. G' band intensity is normalized with $G'_{3DB}$ band of HOPG for easy comparison.

This can also be seen in Table 1 as continuous decrease in the FWHM of D and G bands implying a decrease in defect intensity with increasing $CH_4$ concentration with fixed Ar and $H_2$ contents. Further, except for 5:3:1, the samples grown at feedstock compositions of 1:3:0, 1:0:3, 1:3:1 and 1:1:3 are found to have lower defects in comparison to the other compositions as indicated in Table 1. Thus, it may be inferred that the dilution of $CH_4$ with about 75% of Ar or $H_2$ is optimal to achieve lower defect concentration in their respective morphology.

The line shape, width and position of G' band can be used to probe the number of layers, their structural defects and doping.[32] Fig. 7 shows the Raman spectra of G' band for a few selected films. The intensity of these spectra is normalized with $G'_{3DB}$ band of highly oriented pyrolytic graphite (HOPG) for the sake of clarity. A single Lorentzian peak is found to be the best fit for the G' band as illustrated in Fig. 7. The FWHM values varying from 69.8 to 88.4 $cm^{-1}$ for these samples indicate that these samples contain more than five layers of graphene.[32] However, the single Lorentzian profile signifies that the samples contain stacking fault disorder.[26] The films grown under different compositions also show a turbostratic stacking fault since the G' band could be fitted with single Lorentzian profile for all the samples. This is in support of the observation by HRTEM analysis as discussed earlier. On the contrary, G' band of multilayer HOPG film is composed of G'$_{3DA}$ and G'$_{3DB}$ bands as shown in Fig. 7. The position of G' band of NGSs with respect to G'$_{3DA}$ band of HOPG, is also blue shifted indicating the presence of a few layer graphene.

Table 1: The extracted parameters from Raman analysis for different nanographitic structures grown on SiO$_2$/Si. The topography of the nanostructures are classified as isolated nanoclusters (I.N), vertical (V) and planar (P) structures as given in morphology column. The symbol ' * ' in column '*h*' represents thickness measured by AFM and others values are measured by SEM and *d* represents a discontinuous film.

| Sample | Dilution (%) | morphology | h (nm) | Peak position (cm$^{-1}$) | | | FWHM (cm$^{-1}$) | | |
|---|---|---|---|---|---|---|---|---|---|
| | | | | D | G | G' | D | G | G' |
| | | **Case 1** | **CH$_4$ dilution with Ar** | | | | | | |
| 1:0:0 | 0 | I.N | 12.1* | 1355.4 | 1592.4 | 2703.5 | 57.9 | 57.3 | 109.0 |
| 1:1:0 | 50 | V | 27.8 | 1353.8 | 1587.5 | 2702.2 | 41.6 | 41.5 | 76.8 |
| 1:3:0 | 75 | V | 80.5 | 1353.9 | 1584.7 | 2702.4 | 39.6 | 34.8 | 65.9 |
| 1:5:0 | 83 | V | 136.8 | 1355.9 | 1586.7 | 2704.1 | 39.7 | 36.7 | 69.9 |
| 1:7:0 | 88 | V | 164.9 | 1353.1 | 1584.2 | 2701.5 | 39.3 | 36.1 | 68.6 |
| | | **Case 2** | **CH$_4$ dilution with H$_2$** | | | | | | |
| 1:0:0 | 0 | I.N | 12.1* | 1355.4 | 1592.4 | 2703.5 | 57.9 | 57.3 | 109.0 |
| 1:0:1 | 50 | P | 4.7* | 1356.2 | 1590.1 | 2707.2 | 43.2 | 35.5 | 77.1 |
| 1:0:3 | 75 | P | 5.2* | 1354.5 | 1590.7 | 2705.2 | 37.7 | 34.9 | 69.8 |
| 1:0:5 | 83 | P | 4.3* | 1354.2 | 1590.2 | 2702.9 | 46.8 | 45.2 | 88.4 |
| 1:0:7 | 88 | I.N | -d- | 1352.7 | 1593.0 | 2699.2 | 48.6 | 50.4 | 94.5 |
| | | **Case 3a.** | **CH$_4$ dilution with Ar and fixed H$_2$** | | | | | | |
| 1:0:1 | 50 | I.N | 4.7* | 1356.2 | 1590.1 | 2707.2 | 43.2 | 35.5 | 77.1 |
| 1:1:1 | 67 | I.N | <10 | 1354.2 | 1587.9 | 2701.3 | 41.8 | 40.5 | 75.4 |
| 1:3:1 | 80 | V | 117.5 | 1355.0 | 1585.9 | 2703.9 | 41.4 | 37.1 | 73.3 |
| 1:5:1 | 86 | V | 147.5 | 1352.8 | 1586.4 | 2698.9 | 43.2 | 42.8 | 79.4 |
| 1:7:1 | 89 | V | 160 | 1352.7 | 1585.1 | 2699.8 | 41.7 | 38.2 | 74.3 |
| | | **Case 3b.** | **CH$_4$ dilution with H$_2$ and fixed Ar flow** | | | | | | |
| 1:1:0 | 50 | P | 27.8 | 1353.8 | 1587.5 | 2702.2 | 41.6 | 41.5 | 76.8 |
| 1:1:1 | 67 | P | <10 | 1354.2 | 1587.9 | 2701.3 | 41.8 | 40.5 | 75.4 |
| 1:1:3 | 80 | P | -do- | 1355.8 | 1588.0 | 2708.7 | 41.0 | 34.2 | 72.5 |
| 1:1:5 | 86 | P | -do- | 1353.6 | 1588.8 | 2701.2 | 39.0 | 37.0 | 75.2 |
| 1:3:3 | 86 | I.N/P | -do- | 1355.1 | 1587.2 | 2704.7 | 36.9 | 34.5 | 68.0 |
| | | **Case 3c.** | **Increase of CH$_4$ concentration with fixed Ar and H$_2$** | | | | | | |
| 1:3:1 | 80 | V | 117.6 | 1355.0 | 1585.9 | 2703.9 | 41.4 | 37.1 | 73.3 |
| 3:3:1 | 57 | I.N | 72.4 | 1353.9 | 1585.5 | 2702.1 | 39.0 | 35.8 | 64.6 |
| 5:3:1 | 44 | P | 17.5 | 1355.3 | 1584.8 | 2707.4 | 38.2 | 29.3 | 69.7 |

The planar structures, that are grown under hydrogen dominant feedstock compositions, exhibit a upshift in G band frequency       (upto ~1593 cm$^{-1}$) when compared to the vertical structures as shown in Table 1. The upshift in G band can be attributed to the substrate induced doping whereas free standing vertical graphene has less influence from substrates and hence downshift in G band is observed. The $I_{G'}/I_G$ ratio is also decreased on FPNG in comparison to vertical structures. These results are consistent with the influence of substrate on the Raman spectroscopy of graphene surface.[33] As can be seen from Table 1, the FWHM of the D, G and G' bands clearly decreases upon dilution with Ar or $H_2$ or Ar+$H_2$ and it proves the improvement of crystallinity and reduction of defects with dilution. However, the $I_D/I_G$ ratio is not consistent with FWHM of Raman bands for the studied materials. The vertical structures have more edges whereas the planar structures have more grain boundaries. Hence, the relatively large D-band intensity in both the cases is due to the combination of different types of defects such as stacking fault, hydrogenation, edges, grain boundaries, structural defects from various types of point defects and adsorbed functional groups. In such scenarios where more than two defects contribute, quantification of each of them is very complex and extraction of individual contributions by different types of defects to the Raman band is difficult. Apart from this, the strain in the films is also known to contribute to the change in FWHM or $I_D/I_G$ ratio.[34] However, the present Raman studies reveal that the defect density could be decreased by choosing optimal composition of feed gases.

## Discussion

From the sequence of SEM images, it is observed that dilution of $CH_4$ with Ar always favors growth of vertical graphene structures. Further addition of $H_2$ in fixed proportion (5 sccm) along with Ar produces statistically more stable vertical structures while minimizing the random orientations as evidenced by cross sectional SEM analysis (see Fig. 1 and Fig. 4). On the other hand, the films grown with dilution of $CH_4$ with $H_2$ tend to be planar. However, higher $H_2$ dilution leads to the erosion of initially grown layers into nanoislands over electrically continuous nanographitic film (1:0:5). A further increase in $H_2$ completely wipes out the film and forms a discrete nanographitic clusters over $SiO_2$ surface (1:0:7).

The growth kinetics that dictates formation of different morphological NGSs can be better understood from the radicals produced in plasma under different dilution conditions. A few important chemical reactions between the hydrocarbon and gases used for dilution in the plasma are given below.[24, 35]

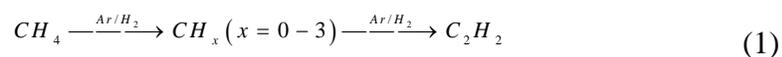

$$CH_4 \xrightarrow{Ar/H_2} CH_x \left( x = 0-3 \right) \xrightarrow{Ar/H_2} C_2H_2 \qquad (1)$$

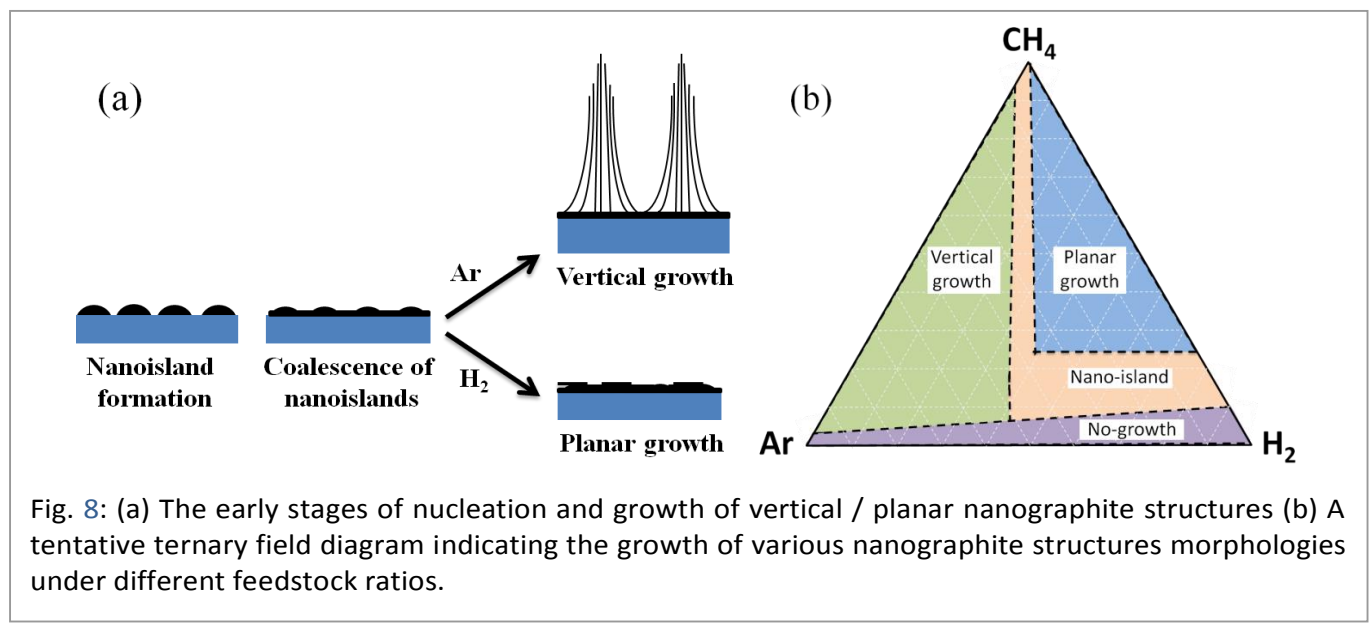

Fig. 8: (a) The early stages of nucleation and growth of vertical / planar nanographite structures (b) A tentative ternary field diagram indicating the growth of various nanographite structures morphologies under different feedstock ratios.

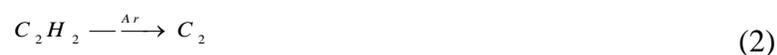

$$(2)$$

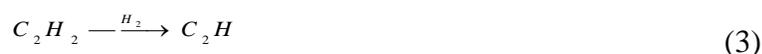

$$(3)$$

In PECVD system, $C_2$ radicals are reported to play a decisive role in determining the surface morphology and growth rate of the films. The availability of $C_2$ radicals can be enhanced by the energetic $Ar+$ ions. However, fine control over the formation of $C_2$ radicals can be achieved by precious control of $H_2$ flow. Such modulation of $C_2$ yields vertical layers of desired aerial density. However the preference of planar morphology in $H_2$ dilution can be understood by the following: Higher dose of $H_2$ increases the density of $C_2H$ radical as shown in eqn. 3 and thereby decreases net density of $C_2$. Further, the available $H^*$ radicals also etch out the growing film.

Fig. 8(a) shows the graphical representation of growth mechanism of NGSs on dielectric substrates. During pre-cleaning procedure, the Ar plasma removes surface contaminations and partial lattice oxygen from $SiO_2$ surface that, creates dangling bonds/active sites on dielectric substrates. The early stage of nucleation is based on direct adsorption of carbon species on these defective $SiO_2$ surface and surface diffusion.[36] This initiates rapid nucleation of carbon nanoislands and formation of a heterogeneous highly defective nanographite buffer layer since the growth rate is extremely high in PECVD. The coalescence of nanoislands and further growth causes stress at nanographitic grain boundaries and whose release favors nucleation of carbon in the vertical direction.[37] Zhao *et al*.[28] reported that the nucleation of vertical growth could start either from buffer layer or from the surface carbon onions. The plasma parameter such as feedstock gas compositions then decides whether the growth of vertical structure to be continued or etched out. If the plasma consists of relatively higher concentration of Ar, it would favor the continued growth of vertical layers. On the other hand, if the erosion is high due to presence of excess $H_2$ concentration, it leads to growth of FPNG films. Furthermore, the plasma parameters such as plasma temperature, radicals' density are entirely different when the $CH_4$ is diluted with $H_2$ instead of Ar. In particular, the temperature gradients at the substrate surface could be considerably lower in the case of $H_2$ dilution, which in turn favor planar growth.[35] Further, the concentration of nascent hydrogen becomes very high in the plasma due to the combined addition of hydrogen while fragmenting $CH_4$ and $H_2$ used for dilution. This high density nascent hydrogen chemically etches out the vertically growing structures.

Based on SEM analysis, an attempt is made to derive a relationship between the morphology and the input gas compositions - $CH_4$, Ar and $H_2$ - towards the growth of NGSs and a ternary field diagram demonstrating the growth regimes is shown in Fig. 8(b). This figure provides an easy understanding of different growth morphologies and corresponding gas compositions. In essence, it is possible to tune morphological features of NGSs with optimal dilution of $CH_4$ with a mixed proportion of Ar and $H_2$.

## Conclusion

We demonstrate the growth and possibility of tunable morphology of nanographite structure (NGSs) on $SiO_2$ substrates using ECR-PECVD by an appropriate composition of feedstock gases. It is found that the dilution of $CH_4$ with Ar and $H_2$ results in vertical and planar structure respectively and also the intermediate structures are always possible with mixed Ar and $H_2$ dilution. The growth mechanism for the different morphological NGSs is due to the competitive growth and etching processes. The Raman studies reveal that the dilution of $CH_4$ with carrier gases minimizes the defects density. Hence, this study helps to grow NGSs with desired morphology and structural quality directly on dielectric substrates using a single growth technique. Also, this study would provide a clue to grow such structures using other plasma based CVD techniques.